\documentclass[conference]{IEEEtran}
\IEEEoverridecommandlockouts
\usepackage{cite}
\usepackage{amsmath,amssymb,amsfonts}
\usepackage{algorithmic}
\usepackage{graphicx}
\usepackage{textcomp}
\usepackage{xcolor}
\def\BibTeX{{\rm B\kern-.05em{\sc i\kern-.025em b}\kern-.08em
    T\kern-.1667em\lower.7ex\hbox{E}\kern-.125emX}}

\usepackage{todonotes}
\usepackage{comment}
    
\begin{document}

\title{Supporting the Adoption of Privacy-Enhancing Technologies through Requirements Engineering\thanks{This research was not sponsored by any of the private companies with which the authors have professional affiliations or engagements. The views, analyses, and conclusions expressed in this paper are solely those of the authors.}}


\author{
    \IEEEauthorblockN{Oleksandr Kosenkov\IEEEauthorrefmark{1}\IEEEauthorrefmark{2}, Vadym Honcharenko\IEEEauthorrefmark{3}, Abhinava Singh\IEEEauthorrefmark{4}, Volodymyr Spirin\IEEEauthorrefmark{5}.  Danica Vranjanin\IEEEauthorrefmark{6}}

    \\
    \IEEEauthorblockA{\IEEEauthorrefmark{1}Blekinge Institute of Technology, Karlskrona, Sweden,
    oleksandr.kosenkov@bth.se}
    \IEEEauthorblockN{\IEEEauthorrefmark{2}fortiss GmbH, Munich, Germany}

    \IEEEauthorblockN{\IEEEauthorrefmark{3}Google, Warsaw, Poland}
    \IEEEauthorblockN{\IEEEauthorrefmark{4}MAYA Data Privacy Limited, Dublin, Ireland}
    \IEEEauthorblockN{\IEEEauthorrefmark{5}adidas AG, Herzogenaurach, Germany}
    \IEEEauthorblockN{\IEEEauthorrefmark{6}JAGGAER LLC, Durham, United States}
}

\maketitle

\begin{abstract}
In recent decades, privacy-enhancing technologies (PETs) have been recognized as a means of meeting regulatory and user privacy requirements in software systems that process personal data. Despite substantial research efforts, support from regulators, contributions by large technology companies such as Google and Microsoft, and growing interest among software practitioners, the practical adoption of PETs remains limited. Existing research consistently identifies recurring challenges to PETs’ adoption in SE, such as technical complexity and insufficient training. Despite ongoing research efforts, these challenges largely remain unresolved in practice.

In this industrial challenge paper, we apply a practical, requirements engineering (RE)-driven perspective to examine challenges to PET adoption across multiple stakeholder groups (PET developers, integrators, and adopters) as well as across different disciplinary perspectives (engineering, law, and business).

We argue that RE can facilitate the adoption of PETs by systematically addressing each of the complementary engineering, business, and legal viewpoints on privacy. Neglecting challenges in any of these viewpoints (e.g., the impact of PETs on software architecture, their business implications, and their contribution to regulatory compliance) can increase the impediments or even lead to implementation failure. In practice, explicit specification of these viewpoints within RE can enable meaningful coordination among stakeholders to more effectively realize the benefits of PETs in software engineering.
\end{abstract}


\section{Introduction}\label{sec:intro}
Privacy is increasingly recognized as one of the important non-functional requirements in software engineering (SE)~\cite{yu2002designing}. The continuous evolution of privacy regulations, together with the increasing use of artificial intelligence (AI), creates pressing challenges for implementing privacy requirements~\cite{solove2025artificial}. One promising privacy engineering approach is the use of privacy-enhancing technologies (PETs)---a class of technical measures that aim at preserving persons' privacy using a variety of techniques. Examples of PETs include homomorphic encryption, which enables computation directly on encrypted data---a concept first introduced in 1978---and secure multi-party computing proposed back in 1982, which allows multiple parties to jointly perform computation while keeping inputs private~\cite{kuhtreiber2025we}.
The adoption of PETs in SE has the potential to improve privacy protection for users by design. While existing studies claim that education and training are the key barriers to the adoption of PETs, we argue that the challenges encountered in practice are considerably more complex.

The complexity of applying PETs in SE stems from the challenges emerging across complementary and interdependent viewpoints on PETs. These challenges cannot be addressed solely through training or better documentation. The traditional discipline-constrained perspective on PETs is insufficient to capture such complexities. Instead, a more pragmatic interdisciplinary perspective is needed to explicitly capture and resolve viewpoint-specific challenges. The complexity of the intersections between different viewpoints can be addressed by requirements engineering (RE) methods, allowing for the modeling of the viewpoints, addressing existing conflicts, and enabling informed and coordinated PET adoption.

In this paper, we argue for the importance of considering engineering, business, and legal viewpoints when adopting PETs. To support this argument, we report the challenges in each of the viewpoints in Section \ref{sec:challenges}. These challenges are primarily derived from the practical experience of the authors with privacy reviews of software systems, PET selection and implementation, implementation of privacy compliance in software products, application of the privacy by design principle in legacy systems, execution of privacy engineering programmes, and third-party privacy risk management. We bring together different perspectives on PETs, such as developers, providers, and adopters, along with the scientific perspective. To identify the reported practical challenges, we used existing scientific studies on this topic (reported in Section \ref{sec:relatedWork}) as a starting point for discussion among the authors and further distilled insights from practice. Finally, in Section \ref{sec:vision}, we propose our vision of how RE can support the adoption of PETs in practice.

\section{Background}\label{sec:background}
In a broad sense, \textit{privacy} can be defined as the right of a person to maintain a private life and personal sphere free from unjustified interference\cite{masur2025comparative}. The ways in which privacy is understood, perceived, and enacted vary across cultural, social, political, economic, and technological contexts. From a technological perspective, this right is supported by \textit{data privacy}, which concerns individuals' ability to control how their personal data is processed.
In this context, \textit{privacy-enhancing technologies (PETs)} are technical measures aimed at enhancing data protection and privacy. PETs may be provided as standalone solutions for end users (e.g., privacy-focused web browsers, secure encrypted email~\cite{caulfield2016adoption}), or as software system components and libraries. The notion of PETs covers a broad range of techniques and technologies~\cite{heurix2015taxonomy} including k-anonymity, differential privacy, homomorphic encryption, and secure multi-party computation. While some PETs can be used to fully anonymize the data, others preserve persons' privacy while processing personally identifiable data. In this paper, we consider PETs on a general level and generalize their purpose as mitigation of risks to data privacy.

\section{Related work}\label{sec:relatedWork}
To explore challenges to the adoption of PETs in practice, we first reviewed prior studies reporting the impact of PETs, challenges to PETs implementation, and proposed solutions to the challenges. Next, we summarize the studies that served as a starting point for our analysis of PET adoption.

\textit{PETs Impact}
Existing secondary studies suggest that PET adoption can affect both technical and business strategy negatively~\cite{thiesse2007assessing}. Empirical studies indicate that PETs have both positive and negative effects on work processes, employee workload, communication patterns, and work completion times. Importantly, these impacts vary across stakeholders and depend on the nature of their work~\cite{gan2019privacy}. These results emphasize that the positive effect of PETs cannot be taken for granted, and their multifaceted effects should be thoroughly analyzed.

\textit{Challenges to PET adoption}
Many studies, including secondary studies, have explored the challenges to PET adoption. We provide an overview and categorize them into groups next.

Challenges related to \textit{PETs complexity}~\cite{zoll2021privacy,klymenko2025supporting,lohmoller2026between} are concerned with the technical side of PETs application (e.g., configuring PET parameters), as well as the complexity of their use from other perspectives. These include their black box nature~\cite{lohmoller2026between}, insufficient understanding of limitations~\cite{lohmoller2026between}, economic risks~\cite{zoll2021privacy}, and residual regulatory uncertainty about whether PETs address all the applicable legal demands~\cite{polonetsky2021review}.

\textit{Knowledge-related challenges} stem from the complexity and include insufficient knowledge~\cite{boteju2023sok,klymenko2025supporting}, or education~\cite{lohmoller2026between}, challenges in selecting appropriate PETs~\cite{klymenko2025supporting}, need for cross-functional interaction, and need to address the terminological differences across the involved functions (e.g., legal and engineering)~\cite{polonetsky2021review,lohmoller2026between}.

Prior work also points to the challenges related to \textit{the current state of PETs development}, including limited maturity~\cite{zoll2021privacy}, maintenance concerns~\cite{zoll2021privacy}, usability issues~\cite{boteju2023sok,lohmoller2026between}, and the need for substantial customization~\cite{lohmoller2026between}. All these require additional resources, including training, as pointed out above.

The aforementioned challenges are complemented by multiple \textit{incentive-related challenges}. Studies point to PET optionality~\cite{klymenko2025supporting}, overall low incentives for adoption~\cite{lohmoller2026between} including lack of regulatory guidance~\cite{zoll2021privacy,lohmoller2026between}, insufficient user demand and consumer readiness~\cite{zoll2021privacy}. There are also a number of organizational challenges that impact incentives such as low privacy priority~\cite{lohmoller2026between,boteju2023sok}, unclear or shifting responsibilities for privacy\cite{polonetsky2021review,boteju2023sok}, limited time and resources~\cite{klymenko2025supporting}, cost-effectiveness of adoption~\cite{klymenko2025supporting,zoll2021privacy,lohmoller2026between}, lack of understanding of a clear relative advantage~\cite{zoll2021privacy} or ways to measure it~\cite{klymenko2025supporting,lohmoller2026between}.

Challenges also emerge from the \textit{state of technologies and processes into which PETs are integrated}. These include inadequate maturity of the software development life cycle (SDLC) to support the integration of PETs (e.g., privacy is not considered from the outset, difficulties in including PETs into system design~\cite{boteju2023sok}), considerable architectural level changes~\cite{boteju2023sok}, need for significant workarounds~\cite{lohmoller2026between},
compatibility~\cite{zoll2021privacy} with legacy systems~\cite{klymenko2025supporting} and existing architectures~\cite{lohmoller2026between}, complications and/or changes in workflows and processes~\cite{lohmoller2026between}, data~\cite{lohmoller2026between} and algorithms~\cite{boteju2023sok}.

\textit{Solutions} for improving PET adoption include organizational measures, such as enhancing awareness, knowledge, and education~\cite{boteju2023sok,klymenko2025supporting,lohmoller2026between}, improved tool support~\cite{lohmoller2026between,klymenko2025supporting}, and improving privacy-related practices across the SDLC~\cite{boteju2023sok}, mapping PETs and the requirements~\cite{klymenko2025supporting}.
In addition, a number of policy-level measures were suggested, such as standardization~\cite{klymenko2025supporting}, increased involvement of regulatory authorities~\cite{klymenko2025supporting}, development of financial incentives~\cite{klymenko2025supporting}, adoption by tech leaders~\cite{klymenko2025supporting}, and addressing the academia-practice gap to facilitate the adoption of research results~\cite{lohmoller2026between}.

\paragraph*{Summary}
A close examination of recent findings reveals that many of the reported challenges were already identified in much earlier studies. In 2011, \cite{borking2011adopting} highlighted the same challenges of complexity, business case identification, usability, and the combination of engineering and legal knowledge. Many corresponding solutions have been implemented over the past years. For example, the Dutch Data Protection Authority was actively promoting the application of PETs up to 2002, and the European Commission issued a Communication on promoting data protection by PETs in 2007~\cite{borking2011adopting}. Notably, \cite{klymenko2025supporting} identified existing solutions for each identified solution category. This raises the question of why the barriers to PET adoption persist despite the abundance of proposed solutions.

While existing research consistently emphasizes education and awareness as the key enablers of PET adoption, the persistence of adoption barriers suggests that these measures alone may be insufficient. We therefore suggest paying attention to the need for interdisciplinary communications to address conflicts and misalignment between different stakeholders, which was only briefly touched upon in some of the studies on PETs~\cite{lohmoller2026between,borking2011adopting,klymenko2025supporting,van2024digital}. For example, ~\cite{lohmoller2026between} suggested coordination across technical, educational, and organizational dimensions, and ~\cite{pelkola2012framework} proposed the establishment of multidisciplinary steering committees for PETs adoption; however, neither study provided concrete approaches to implementation. Such an approach is, however, more prominent in the studies on regulatory RE and regulatory compliance in SE~\cite{klymenko2022understanding,kosenkov2025systematic}. Such studies also empirically identified the differences across the viewpoints. For example, while engineers are primarily interested in requirements-to-implementation traceability, legal experts demand traceability from the text of regulations to implementation \cite{kosenkov2025privacy}. We argue that accounting for different viewpoints and analyzing challenges through these viewpoints provides a more effective way to understand and facilitate PET adoption. Next, we provide an overview of the challenges associated with three distinct viewpoints.

\section{Viewpoints on PETs and their adoption}\label{sec:challenges}
In this section, we intentionally focus on three viewpoints that directly affect the adoption of PETs in SE and that the authors have encountered in practice. Nevertheless, we acknowledge that it is important to account for additional viewpoints on PETs and privacy, such as those of users.
It is noteworthy that in practice, many of the challenges intertwine across the viewpoints. For example, the legal challenge of the relativity of data anonymity has implications for the engineering challenge of achieving enterprise-wide data visibility. However, we discuss these challenges in isolation and without prioritization due to space limitations.

\subsection{Engineering viewpoint}
The engineering viewpoint reflects how software engineering and information technology professionals approach the implementation and maintenance of software systems and required privacy-enhancing technical measures. The following challenges to PET adoption are characteristic of this viewpoint.

\paragraph{Need for enterprise-wide data visibility}
In SE research and practice, privacy is often approached as a challenge of processing personal data within a single software system. In fact, privacy is concerned with the processing of personal data across multiple software systems used in an organization. Effectively, it is challenging to provide visibility into all data processing activities and data-related demands across all software systems. Because many PETs can irreversibly transform data, limited visibility into data processing may cause companies to reject PETs to avoid unexpected consequences in downstream systems and processes.

\paragraph{Assessment of PETs deployment options}
Organizations can integrate PETs into their software and IT architecture in various ways. However, analyzing and selecting PETs deployment options depends on both the purposes for PETs usage and their impact on the architecture. Simultaneously, PET adoption is usually driven by regulatory non-functional requirements, rather than typical functional or non-functional requirements. Consequently, informed decision-making demands a combination of specialized legal and technical knowledge. For example, when considering the deployment of PETs in AI-enabled systems, it is important to take into account data memorization and algorithmic shadow~\cite{li2022algorithmic,hutson2025forget}, which are particularly complicated to address in AI technologies. Techniques such as machine unlearning can reduce the influence of personal data on model outputs. However, it remains an open question whether these techniques fully comply with data protection laws. While a model's weights can be adjusted to lessen the impact of a specific record (e.g., through fisher-scrubbing), the model may still produce inferences based on the removed or altered data. This situation may not adequately address data subjects' rights to data deletion or rectification~\cite{hutson2025forget}. 

\paragraph{Impact on architectural practices}
Traditional software architecture specification techniques often do not capture all information required for privacy engineering (e.g., detailed data flows). As a result, engineers may need to maintain a separate architectural privacy viewpoint or adopt privacy modeling techniques (e.g., privacy threat modeling) to support PETs implementation.
In addition, the involvement of a legal expert may be necessary to guarantee that compliance is achieved and sufficient evidence of implementation is available. Especially when PETs implementation is driven by the legal or privacy experts, it can be essential to establish engineering-legal communication to compensate for the absence of detailed technical knowledge of legal experts and for the absence of legal or privacy knowledge of engineers. Once legal and privacy engineering are aligned on the specific goals of the implementation of PETs, metrics can be established and maintained to evaluate the effectiveness of these privacy-preserving techniques. Additionally, product teams, information security, and marketing teams can be involved when discussing the potential for leveraging PETs within the organization.

\paragraph{Lack of a privacy-focused mindset among engineers}
Another barrier to PET adoption is the limited recognition of the importance of privacy as a non-functional requirement. While a transition to more privacy-focused architectural and development practices is required, engineers tend to focus on functional requirements and features. Engineers may also approach privacy engineering and PETs through the lens of familiar paradigms (e.g., treat data masking as a data type). Here, the challenges stem not from insufficient knowledge of PETs but from insufficient understanding of privacy engineering and recognition of its importance.     Observations from practice suggest that start-ups tend to use PETs because their mindset embeds privacy as one of the essential non-functional requirements from the very beginning.

\paragraph{Data transformation and changes in processing activities}
Many PETs transform personal data in ways that are often irreversible. As a result, PET adoption introduces changes and constraints in processing the transformed data in comparison to conventional data processing. This might require robust data governance practices to preserve the same level of data management (e.g., data mapping might be impacted).
For example, when PETs are applied but complete anonymization is not achieved, organizations may need additional measures to retrieve or delete data in case of a corresponding request from a data subject.

\paragraph{Existing PETs are often case-constrained}
Many PETs were developed to satisfy very specific needs, but only later were they made public for broader use. As a result, identifying suitable use cases and adapting PETs correspondingly can be challenging. Here, the development of PETs for specific cases or industries and the improvement of PETs documentation can be suitable measures on the side of PET developers.

\paragraph{Limited technical incentives}
Beyond privacy protection and regulatory compliance, there are only a few direct technical benefits of the PETs in SE. Therefore, it is important to identify concrete benefits for engineering teams (e.g., minimization of privacy controls implementation efforts) to motivate engineers to learn technologies and implement them.

\paragraph{Maintenance and configurability of PETs}
Once PETs are integrated into a software system, they must be maintained and, among other things, reconfigured as data processing evolves. Maintaining PETs is a special challenge because of the shared responsibility between developers and legal experts (as discussed in the legal viewpoint section) and the need to consider all the applicable requirements during the maintenance (e.g., regulatory requirements, user requirements).

\paragraph{Heterogeneous techniques under the PETs umbrella}
PETs are often presented as a single type of privacy technique (and often perceived that way by non-technical roles). In reality, the term encompasses techniques that are hard to compare from the technological perspective~\cite{heurix2015taxonomy}. This complicates the communication between engineering and non-technical roles and does not allow for effective selection of PETs by focusing on particular properties. Existing PET taxonomies (e.g.,~\cite{heurix2015taxonomy}) provide useful classification, but do not effectively support differentiation and selection of PET techniques in practice.

\subsection{Business viewpoint}
The business viewpoint focuses on the operational and business aspects of PETs implementation and is usually adopted by product owners, project managers, and other business stakeholders. This viewpoint is focused on ensuring that the business case for PET adoption is in place, which can require accounting for different effects of PETs.

\paragraph{Impact on the value of data}
PETs affect the utility of data for business purposes, including data analytics and monetization. Consequently, organizations need to re-evaluate existing business models, particularly when business models and processes were not designed with privacy in mind from the outset. Therefore, the impact of PETs should be considered to balance privacy and business requirements.

\paragraph{Change of business processes}
Another practical challenge mentioned in existing studies and confirmed by our experience is the need to change business processes in case PETs have a significant impact on the processed personal data. When re-engineering of business processes and further operational adaptation are not taken into account during PET adoption, organizations may incur additional expenses. Even when organizations can cut costs due to PETs, this should be considered in the RE process.

\paragraph{Stakeholders on multiple levels of organization}
Understanding the organizational implications of PET adoption is hampered by the multifaceted implications of PETs. This demands the involvement of a multitude of stakeholders across different functions and levels in the organization. For example, a clear business or legal value does not guarantee easy adoption because of the potential technological constraints. Consequently, identifying and aligning all these impacts requires additional resources and methods.

\paragraph{Realizing reputational benefits}
Reputational benefits associated with PETs appear to be more significant for large organizations than small and medium enterprises, which may derive fewer competitive advantages from their use. In addition, the application of PETs can be hard to demonstrate to customers and may require raising overall awareness among customers about privacy and PETs.

\paragraph{Impact on privacy compliance risks}
From a business perspective, PETs application does not necessarily and proportionally minimize compliance risks. Hence, it can be difficult to quantify the value of PETs, especially against the implementation expenses. Further, when PET adoption is narrowly focused on compliance, it may amplify the risks to privacy in a broader societal sense and even be seen as “privacy washing”, thus undermining consumers’ trust~\cite{murphy2026protecting}.

\subsection{Legal viewpoint}
The legal viewpoint is concerned with implementing and demonstrating compliance with applicable privacy regulations. In a broad sense, it can also include compliance, i.e., a broad set of measures for implementing compliance (e.g., conducting training, raising awareness).

\paragraph{PETs do not guarantee privacy compliance}
The application of PETs in many cases does not automatically result in privacy compliance as required per regulations~\cite{ico2023pets}. PETs can help to demonstrate compliance with the principles of data minimization, privacy by design, and by default. However, for establishing privacy compliance, PETs need to be assessed along with other technical and organizational measures. This way, the legal viewpoint depends on the engineers and other roles implementing privacy compliance controls to have up-to-date information about the corresponding technical measures for assessment of PETs within a broader compliance context.

\paragraph{Selection of complementary organizational measures}
Like other technical privacy measures, PETs require complementary organizational measures. The identification of such organizational measures depends on the circumstances of the specific processing activity. This requires that legal experts are qualified to take into account PETs as an additional factor.

\paragraph{Privacy and compliance are moving targets}
Regulations are intentionally abstract and vague to provide flexibility in their implementation~\cite{breaux2007systematic,kosenkov2024developing}. Moreover, the interpretation of privacy and the required appropriate controls also evolve over time, making privacy a moving target. As legal experts are responsible for interpreting regulatory norms in each specific case, it can be hard to communicate such evolving and intricate requirements to stakeholders belonging to other viewpoints and to position PETs within the context of other controls, and overall evolving interpretation of privacy.

\paragraph{Complex relationships between roles within shared responsibility}
Shared responsibility is a fundamental characteristic of the modern technological landscape because organizations depend on their partners in the fulfillment of compliance and security requirements. The complexity of PETs often causes adopters and users to rely heavily on technology developers and providers to fulfill compliance and privacy duties. Legal practice in some cases can qualify technology providers as joint data controllers, increasing their responsibility (e.g., in the Russmedia case \cite{cjeu2025russmedia}). As regulatory compliance remains the main motivation for PET adoption, the capacity of PET developers and providers to support adopters and users in the fulfillment of regulatory compliance requirements, thereby sharing responsibility for compliance, is essential. Clear differentiation of duties within shared responsibility should also help engineering roles know their duties clearly and define responsibility for PETs in companies.

\paragraph{Relative notion of anonymity}
Existing regulations define data anonymity relative to the likelihood that an individual can be identified using all the means likely reasonably to be used  (e.g., all the data possessed by a particular organization and/or publicly available data)~\cite{iapp2023anonymization}. This creates challenges in identifying the required degree of anonymization and subsequently the application of PETs. It also means that some PETs may be incapable of fully satisfying such requirements~\cite{wp29}. As a result, legal experts need insights into existing data architecture (available data and processing activities) or the technical state of the art.

\paragraph{Differences across jurisdictions and regulations}
Despite the encouragement of PETs by regulators, there are still different approaches by regulators to PETs in different jurisdictions (e.g., need for consent for synthetic data generation~\cite{pilgram2025protecting}). Moreover, PETs may serve as measures for different regulations (e.g., privacy and security). Thus, it could be challenging to select PETs that would satisfy the requirements across different jurisdictions and regulations (e.g., avoid conflict between the application of PETs and fairness requirements~\cite{calvi2024unfair}). While harmonization of regulations is possible to a limited degree only, more systematic regulatory requirements engineering with a better inclusion of legal experts having insights into such differences can be a way to address this challenge.

\paragraph{Measuring residual compliance and risk}
As PETs that do not provide complete anonymization do not guarantee regulatory compliance, it is essential to be able to measure residual risks and identify additional measures required after the application of PETs. However, this task introduces additional work and complexity for legal experts demanding knowledge of PETs (e.g., integration of PETs into threat modeling and risk management activities).

\paragraph*{Summary} Addressing these challenges within each of the viewpoints is challenging, and the interplay between the viewpoints makes it essential to use special approaches to prioritize and resolve challenges across viewpoints.

\section{Practical Lessons Learned on PET adoption}
However, viewpoint interaction is essential not only to resolve challenges, but also to realize benefits. In the authors’ experience, one of the main benefits of adopting PETs in SE is the enhancement of software trustworthiness and user acceptance. However, realizing these benefits largely depends on the ability of development teams to integrate PETs early and effectively. To maximize their added value from a users’ perspective, PETs should be viewed as one of architectural elements that contribute to privacy as software quality. They should be consistent with, and complementary to other privacy controls that are visible and meaningful to users.

Ideally, PETs should be integrated into software products by design and from the outset of their lifecycle. This helps avoid technical and organizational overhead associated with later adoption. Early integration also maximizes the potential positive impact on trustworthiness, whereas adopting PETs after a product has been introduced to the market can have less noticeable benefits.

Treating PETs as an integral part of the baseline privacy by design is particularly important for complex and ubiquitous software products, such as AI-enabled systems. The absence of visible privacy safeguards, including PETs, may negatively affect user’s perception of the overall product quality.

Effective coordination among viewpoints and stakeholders can be one of the key prerequisites for successfully embedding PETs into system’s privacy architecture effectively and maximizing the benefits of their adoption.

\section{Vision for the Contribution of RE to PET Adoption}\label{sec:vision}
One of the core tasks of RE is to align the needs and expectations of different stakeholders to identify solutions that best satisfy these needs. In the context of PET adoption, this task is particularly complicated because it requires consideration of technical constraints alongside complex legal and business concerns. Although viewpoint-based RE approaches have not gained substantial traction since their emergence in the early 2000s, we argue that RE is crucial for addressing the challenges to PET adoption across multiple viewpoints.

First, RE stakeholder modeling approaches can be used to systematically identify the relevant stakeholders and viewpoints. Such approaches can also raise awareness about the complementary nature of these viewpoints.

Second, practitioners adopting PETs require explicit and specific information about the demands, constraints, and challenges associated with other viewpoints. RE methods that are applicable across viewpoints, such as general goal-based or privacy-specific approaches~\cite{kosenkov2026towards}, constraint modeling~\cite{torre2021modeling}, could be valuable for capturing and communicating these requirements stemming from different viewpoints separately. Furthermore, PET adoption also involves a legal viewpoint characterized by substantial knowledge intensity~\cite{kosenkov2025systematic}. RE can help structure this complex knowledge and facilitate engineering-legal interaction. This includes both making legal knowledge more understandable to engineers and making engineering knowledge more accessible to legal experts.

Third, many of the most significant barriers to PET adoption stem from the intersections and collisions between the viewpoints. In this context, the existence of specifications and basic traceability is instrumental for resolving such conflicts.

Fourth, taken together, the RE contributions outlined above enable effective coordination among viewpoints. This encompasses structured and informed discussions, the ability to make decisions, and the capacity to keep PETs relevant throughout regulatory, business, and technological changes.

Overall, the case of PET adoption illustrates the central role of RE in balancing and coordinating the interests and goals of software engineering, legal, and business stakeholders during the development of software-intensive products and services.

\section{Conclusion}
As observed in practice, PET adoption involves at least three complementary viewpoints, each associated with challenges. We argue that requirements engineering can play a pivotal role in addressing the multi-viewpoint challenges in PET adoption.

First, RE can help manage heterogeneous viewpoints by modeling them and the corresponding challenges. Second, RE is instrumental for domain and knowledge engineering of the complex legal viewpoint. Third, RE can be helpful to address the interdependencies between the viewpoints. Fourth, and most important, RE can facilitate the coordination and decision-making to drive more efficient and effective adoption.

While in this paper we mainly discussed the adoption of PETs, we suggest that applying requirements engineering techniques would also be useful for the development of future PETs, which would address the challenges and satisfy the demands of the involved viewpoints.

We hope that the insights provided in this paper will be useful to researchers and practitioners alike, as well as to both the requirements engineering and privacy engineering communities. We further call for industry-focused empirical research that explores PET and privacy-requirements challenges ``in the wild'' to ensure that research findings and recommendations are better aligned with practical needs.

\bibliographystyle{ieeetr}
\bibliography{bibliography.bib}

\end{document}